\documentclass[12pt]{article}
\usepackage[pdftex]{color, graphicx}
\usepackage{amsmath, amsfonts, amssymb, mathrsfs}
\usepackage{lscape}
\usepackage{float}
\usepackage{dcolumn}
\usepackage{changepage}
\usepackage{hyperref}
\usepackage{amsthm}
\usepackage{bbm}
\usepackage{amsmath}
\usepackage{setspace}
\usepackage{subcaption}
\usepackage{float}

\date{}

\title{Embedding Positive Process Models into Lognormal Bayesian State Space Frameworks using Moment Matching\vspace{-.5ex}}
\author{Smith, J. W., Johnson, L. R., Thomas, R. Q.}
\author{John W. Smith$^{1,2}$ \and Leah R. Johnson$^{2,4}$ \and R. Quinn Thomas$^{3,4}$}
\date{%
    \small
    $^1$Department of Mathematical Sciences, Montana State University\\
    $^2$Department of Statistics, Virginia Tech\\%
    $^3$Department of Forest Resources and Environmental Conservation, Virginia Tech\\
    $^4$Department of Biological Sciences, Virginia Tech\\
    [0.5ex]%
}
\usepackage{natbib}

\begin{document}

\maketitle
\abstract{In ecology it is common for processes to be bounded based on physical constraints of the system. One common example is the positivity constraint, which applies to phenomena such as duration times, population sizes, and total stock of a system's commodity. In this paper, we propose a novel method for embedding these dynamical systems into a lognormal state space model using an approach based on moment matching. Our method enforces the positivity constraint, allows for embedding of arbitrary mean evolution and variance structure, and has a closed-form Markov transition density which allows for more flexibility in fitting techniques. We discuss two existing lognormal state space models, and examine how they differ from the method presented here. We use 180 synthetic datasets to compare the forecasting performance under model misspecification and assess estimability of precision parameters between our method and existing methods. We find that our models well under misspecification, and that fixing the observation variance both helps to improve estimation of the process variance and improves forecast performance. To test our method on a difficult problem, we compare the predictive performance of two lognormal state space models in predicting Leaf Area Index over a 151 day horizon by embedding a process-based ecosystem model. We find that our moment matching model performs better than its competitor, and is better suited for long predictive horizons.  Overall, our study helps to inform practitioners about the importance of embedding sensible dynamics when using models complex systems to predict out of sample. }

\section{Introduction}\label{sec1}

Process based models are mathematical representations of the evolution of biological or physical systems \citep{Buck-Sorlin2013}. These models are often comprised of systems of ordinary or partial differential equations in time and space, or are discretizations of such systems. Because they encapsulate known or hypothesized mechanisms of physical or biological systems, process based modeling approaches have advantages over empirical or phenomenological modeling approaches, particularly when low data availability limit the ability of empirical models to accurately represent complex processes. As a result, process based models remain common in ecological forecasting applications  \citep{LewisReview}. However process-based modeling approaches have their own set of challenges, particularly the quantification of uncertainty in model dynamics and structure or over-parameterization  \citep{LUO2009}.

Quantifying uncertainty in process based models is one of the most challenging tasks when using them for forecasting applications that involve uncertainty propagation. Uncertainty comes from a wide variety of sources, including but not limited to: process, measurement, initial conditions, driver data, and estimated parameters \citep{DIETZE2018}. Process uncertainty (or process stochasticity) is particularly important to address, as it acknowledges that the modeling framework may contain unknown errors that are best represented stochastically or contain elements that are known to be non-deterministic and that affect the dynamics of the biological process. For this reason, the state space modeling (SSM) framework \citep{DurbinKoopman, Petris, am} has been used frequently in ecological applications  \citep[see][for examples]{THOMAS2017, DOWD200339, dennis06}. State space models provide a flexible framework that is able to handle missing data and partition multiple sources of uncertainty \citep{DIETZE2018}. Many existing fitting methods for ecosystem process based models already estimate initial condition and parameter uncertainty while accounting for observation uncertainty. By adding stochastic elements to the process based model they can be analyzed as state space models.  

One of the nuances of including uncertainty/stochasticity in the process model is the trade-off between complexity and ecological realism. For example, in forest carbon modeling, Gaussian error, with its positive to negative infinity bounds, is commonly assumed for carbon stocks (pools) \citep{THOMAS2017, JIANG2018}, despite the biophysical impossibility of an ecosystem having a negative amount of carbon. In general, biological processes may have well defined lower bounds (and potentially upper bounds) that are not accurately captured by error structures that have support over the entire real line. Consequently, models that have biophysically unrealistic error structures can produce nonsensical predictions as the states approach these well defined bounds, or if the observations have large measurement error which can allow for predictions  with negative values as the modeled states approach the lower bounds. This is especially relevant in forecasting applications, where the uncertainty compounds as the forecast horizon increases. 

A wealth of probability distributions are available for modeling non-negative processes. In particular, the lognormal distribution has a rich history in ecology \citep{dennis_chapter}, and is frequently used in state space modeling of populations \citep[e.g., ][]{BUCKLAND2004, dennis06, Knape2011} and species abundance \citep[e.g, ][]{MAUNDER2015, samu2015}. Two common formulations for lognormal state space models (LN-SSMs) are the stochastic Gompertz SSM \citep{gompertz} and the stochastic Moran-Ricker SSM \citep{ricker, dennis06}. In their simplest forms, these models include assumptions that are disguised by writing the models as Gaussian in log space - namely the assumptions of density dependent variance and systematic bias in the process evolution and observation functions. When these assumptions are not feasible for an application, we need a mechanism to insert more appropriate assumptions about the process evolution, observation, and variance dynamics. To address challenges with incorporating ecologically realistic error structures into state space models, we propose a novel lognormal moment matching approach that allows users to specify the mean and variance of their process evolution and observation models.

Here, we fit our lognormal moment matching models from a Bayesian perspective using Markov Chain Monte Carlo (MCMC) and Particle Markov Chain Monte Carlo \citep[pMCMC;][]{pMCMCAndrieu}. The Bayesian paradigm provides a flexible framework for fitting complex models, and the moment matching approach we introduce here offers a closed form Markov transition density. This closed form transition density provides the option to fit these models using MCMC, while still supporting access to particle filter \citep{cappe, pf_15years_later} methods such as pMCMC. MCMC and pMCMC also provide a rigorous framework for quantifying parameter uncertainty and assessing forecast performance. MCMC methods generate samples from the posterior distribution, allowing practitioners to generate parameter estimates and build empirical density functions for their forecasts \citep{kruger2021}. We can then validate our models by combining out of sample forecast observations with MCMC output and evaluating them with proper scoring rules \citep{strictly_proper}, such as the Continuous Ranked Probability Score  \citep[CRPS;][]{matheson_winkler1976} and the Ignorance score  \citep[IGN;][]{good1952, EvaluatingProbabilisticForecastsUsingInformationTheory}. 

We create four different models using the lognormal moment matching technique that we will present here. The four models are all based off of the Gompertz and Moran-Ricker models \citep{gompertz, dennis06, ricker}, and are embedded (put into the latent process model in a particular way) to have unbiased process evolution. We explore two different variance structures: a density dependent variance \citep{dennis06} for the evolution and observations, and a constant variance for the evolution and observations. First, we discuss interpretations of the Gompertz and Moran-Ricker SSMs  and contrast them with interpretation of the models we present here. Next, we design and conduct simulation studies to compare forecast performance under model mis-specification, and assess estimability of precision parameters. Finally, we embed a two-dimensional process-based ecosystem model based on the DALEC2 model of \cite{BLOOM}, and use it to predict Leaf Area Index (LAI) at a focal forest site (University of Notre Dame Environmental Research Center; UNDE) in Wisconsin, USA. We perform the embeddings in two different ways: once using a biased embedding with a density dependent variance structure, and once using our moment matching embedding with a density dependent variance structure. We design and perform an analysis to test the viability of using a state space modeling framework to predict out of sample LAI, and to identify similarities and differences between our moment matching technique and the more standard biased embedding.

\section{Motivating Examples}

In this section, we motivate the use of biologically realistic error structures in applications by demonstrating problems that may occur when modeling positive processes with error distributions that have mass over the entire real line. In particular we demonstrate how process models produce nonsensical predictions as they approach or exceed well defined biophysical bounds. First, we consider the following toy dynamical system in time:
\begin{align}
X_t \lvert X_{t-1} \sim \text{log} \mathcal{N}\left(\mu^* = \log\left(\frac{X_{t-1}^2}{\sqrt{X_{t-1}^2 + \sigma^2}}\right), \sigma^{2*} = \log\left(1 + \frac{\sigma^2}{X_{t-1} ^2}\right)\right). \label{eq:LNDynSys}
\end{align}
Since the stochastic evolution function is lognormally distributed, the process is positive, i.e. $X_t \in (0, \infty), \forall t = 1, \dots, T$. Further the conditional mean and variance are $\mathbb{E}[X_t \lvert X_{t-1}] = X_{t-1}$ and $\mathbb{V}[X_t \lvert X_{t-1}] = \sigma^2$. 

\begin{figure}[ht!]
\begin{adjustwidth}{1cm}{1cm}
\centering
\includegraphics[scale = .35]{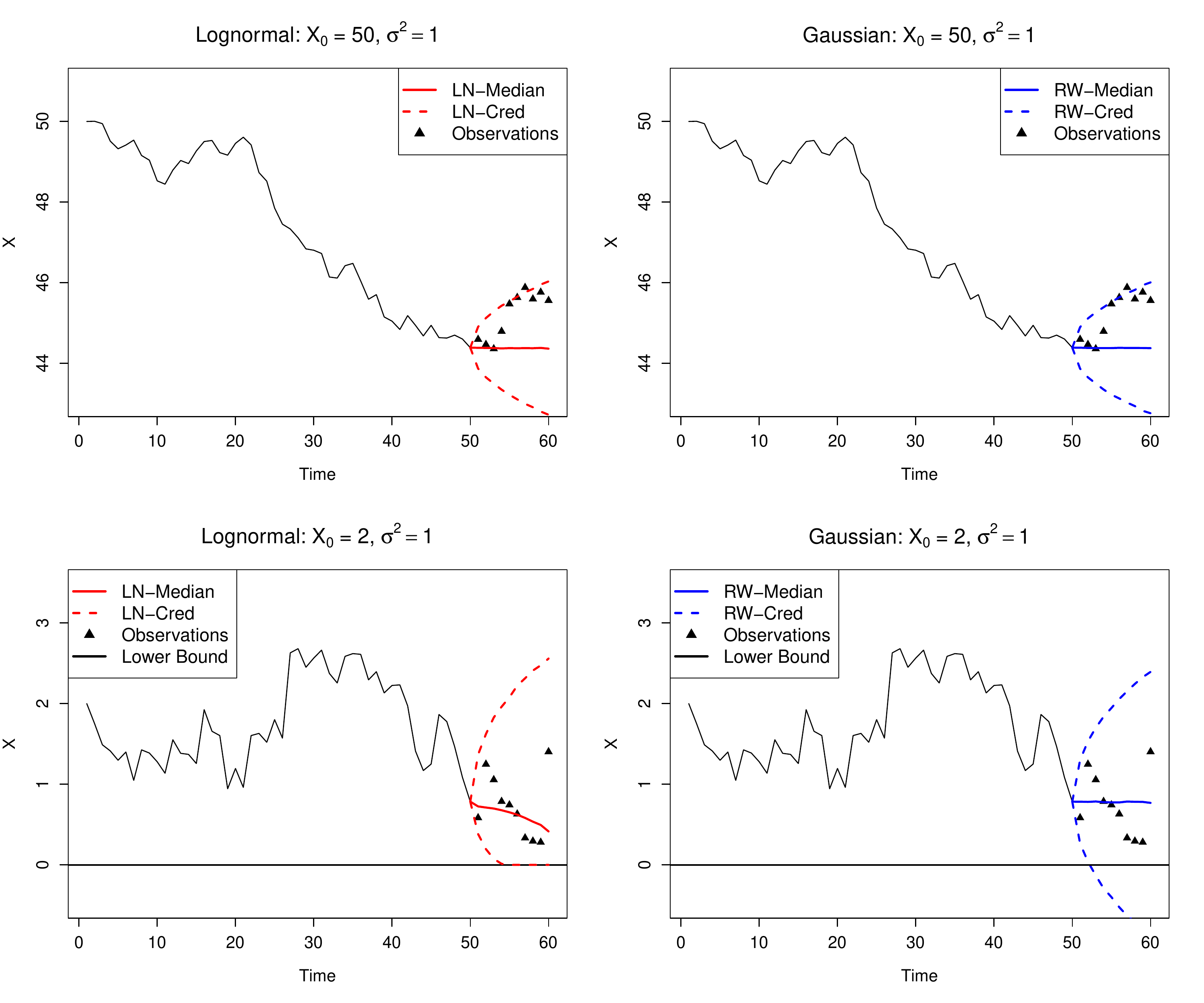}
\caption{Two sample trajectories from the Lognormal dynamical system in Equation \ref{eq:LNDynSys}, the top panels start from $X_0 = 50$ and the bottom panels start from $X_0 = 2$. 95 \% credible intervals for 10 day forecast horizons (dotted lines) are created using the Lognormal equations and the Gaussian equations to generate 10,000 sample prediction trajectories. In the bottom panels, the Lognormal forecasts (bottom left) are bounded below by zero, the biological lower bound. In contrast, the Gaussian random walk forecasts (bottom right) put considerable forecast probability mass below the biological bound.
\label{fig:m_ex1_plot}}
\end{adjustwidth}
\end{figure}

Suppose that the positivity of the dynamical system is ignored and instead the process is modeled as simple a Gaussian random walk:
\begin{align}
X_t \lvert X_{t-1} \sim \mathcal{N}(\mu^* = X_{t-1}, \sigma^{2*} = \sigma^2). 
\end{align}
When the system starts sufficiently far from zero the two models are indistinguishable from each other, producing nearly identical forecast distributions in terms of median estimates and credible intervals (Figure \ref{fig:m_ex1_plot}, top two panels). When the starting value $X_0$ is chosen to be close to the biological lower bound, the differences in forecasts become more apparent. The lognormal model forecast intervals are now asymmetric and bounded below at zero, and quickly reach zero as the forecast horizon increases. The random walk model forecast intervals retain their symmetry and continue to put considerable forecast mass below zero as the forecast horizon increases (Figure \ref{fig:m_ex1_plot}, bottom two panels). 

In this simple example, the negative forecasts are not causing any issues in the process model, thus only producing unrealistic forecasts. To illustrate an example where issues arise in the process model, consider the following dynamical system:
\begin{align}
X_t \lvert X_{t-1} \sim \text{log} \mathcal{N}\left(\mu^* = \log(\lvert X_{t-1}\lvert) - a, \sigma^{2*} = \log\Big(1 + \frac{\sigma^2}{\lvert X_{t-1}\rvert}\Big)^2\right). \label{eq:LNDynSys2}
\end{align}
We note that, formally, absolute values are unnecessary for the evolution of the true model (since $X_t$ is positive for all $t$). However they will become necessary when we attempt to find an analogous Gaussian model with which to forecast. 

Computing the conditional mean and variance of the lognormal distribution with the given values of $\mu^*$ and $\sigma^{2*}$ and using them as the conditional mean and variance of a Gaussian distribution results in:
\begin{align}
X_t \lvert X_{t-1} \sim \mathcal{N}\left(\exp\left(\mu^* + \frac{\sigma^{2*}}{2}\right), \left(\exp(\sigma^{2*}) - 1\right) \exp\left(2\mu^* + \sigma^{2*} \right)\right).
\end{align}

\begin{figure}[ht!]
\begin{adjustwidth}{1cm}{1cm}
\centering
\includegraphics[scale = .39]{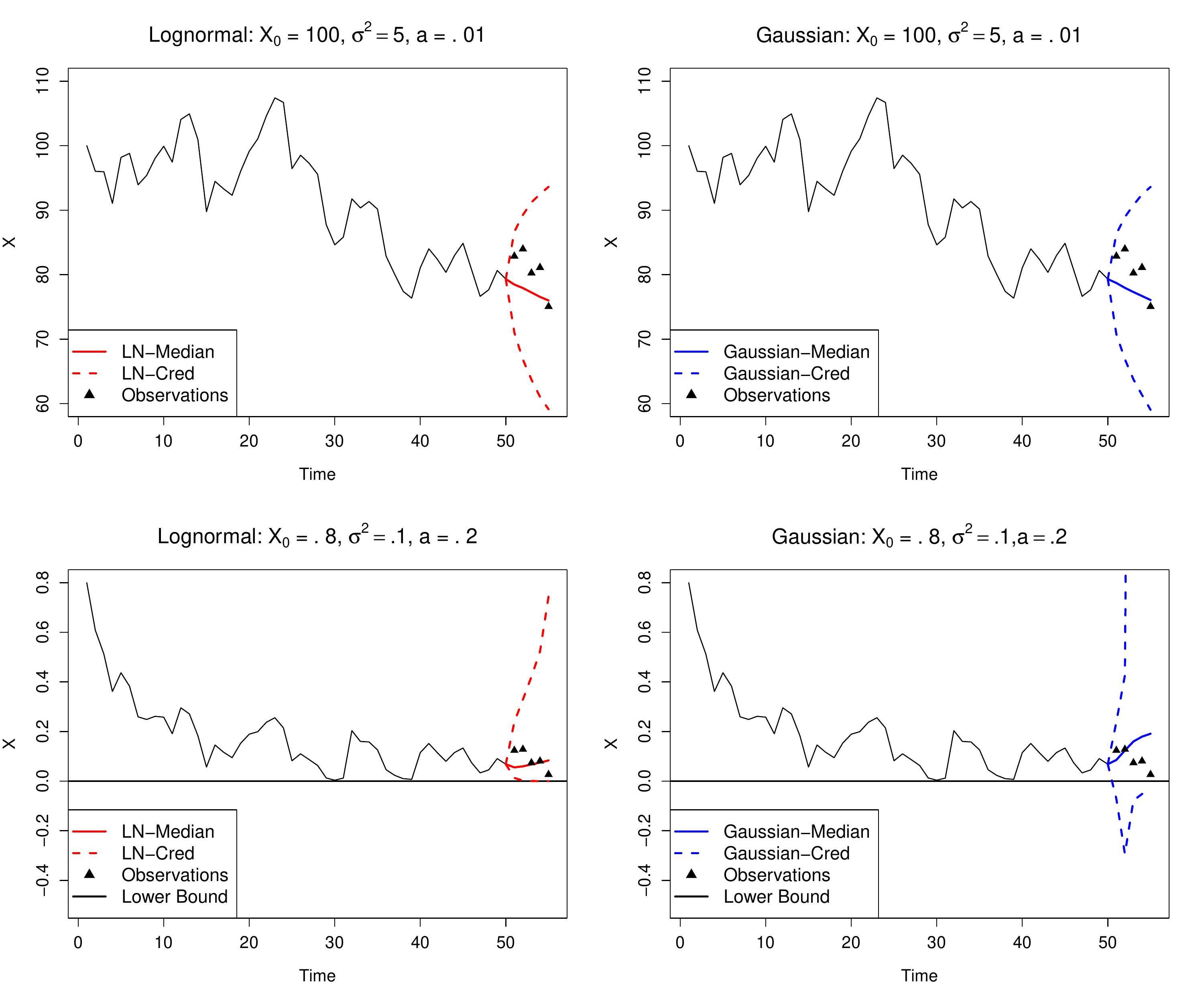}
\caption{Two sample trajectories from the Lognormal dynamical system in Equation \ref{eq:LNDynSys2}, the top panels starting from $X_0 = 50$ and the bottom panels starting from $X_0 = 2$. 95\% credible intervals for 10 day forecast horizons (dotted lines) are created using the Lognormal equations and the Gaussian equations to generate 10,000 sample prediction trajectories. In the bottom panels, the Lognormal forecasts (bottom left) are bounded below by zero, the biological lower bound. In contrast, the Gaussian random walk forecasts (bottom right) put considerable forecast probability mass below the biological bound.}
\label{fig:m_ex2_plot}
\end{adjustwidth}
\end{figure}

For values sufficiently far from zero, the forecasts from the Lognormal model are nearly identical to the forecasts from the Gaussian model (Figure \ref{fig:m_ex2_plot}, top two panels). The Lognormal forecasts (bottom left panel, Figure \ref{fig:m_ex2_plot}) enforce the lower bound, and produce forecast distributions with heavy upper tails. However, the Gaussian forecasts when the system approaches zero (bottom right panel, Figure \ref{fig:m_ex2_plot}) produce forecast distributions with large upper credible bounds that march away from zero, and lower credible bounds that first increase and then move back towards zero. In working on specific applications, predictions like these may be challenging to analyze because it is easy to associate the dynamics with an inadequate process model, when in reality the problem lies in having an error structure that can put probability mass on regions where it should be biologically impossible.

\section{Methods}
\subsection{Lognormal SSMs}

State Space Models, sometimes referred to as Hidden Markov Models (HMMs), are a broad class of models used to track the states of a model (some unobserved process $X_{1:T}$) through a set of observations, $Y_{1:T}$ \citep{DurbinKoopman, Petris}. A state space model has three components -- the state component, the observation component, and additional parameters. The state component consists of an (unobserved) Markov process, $X_{t}$, being moved through time by an evolution function (or process function) $f(X_t \lvert X_{t-1}, \Theta)$. The implication of $X_{1:T}$ following a Markov process is that the past and the future are independent conditional on the state at the current time, $X_t$ \citep{Shumway}. The observation component consists of noisy observations of the latent process, $Y_{1:T}$, that are governed by an observation density function, $g(Y_t \lvert X_t, \Theta)$. These observations are assumed to be independent of one another conditional on the latent states $X_{1:T}$.  The additional parameter component, $\Theta$, contains the parameters that govern the evolution function and observation function.

The Gompertz \citep{gompertz} and Moran-Ricker \citep{ricker} SSMs are simple discrete density-dependent state space models with lognormally distributed process and measurement error \citep{dennis06}. The Gompertz and Moran-Ricker SSMs are popular choices for lognormal state space models because they can be easily transformed into Normal Dynamic Models (NDMs), and can then take advantage of a suite of well studied fitting methods including Kalman filtering \citep{Kalman}, extended Kalman filtering \citep{Julier1997newEO}, and Gibbs sampling \citep{GemanAndGeman, GaussianGibbs}, easing computational difficulties associated with fitting SSMs.  The latent process models for the Gompertz model can be written as:
\begin{align}
X_t &= X_{t-1} \exp (a + b \log(X_{t-1}) + \epsilon_t), \text{ } \epsilon_t \sim \mathcal{N}(0, \phi) \label{eq:gompertz} 
\end{align}
The latent process model for the Moran-Ricker model can be written as:
\begin{align}
X_t &= X_{t-1} \exp (a + b X_{t-1} + \epsilon_t ), \text{ } \epsilon_t \sim \mathcal{N}(0, \phi) \label{eq:MoranRicker}
\end{align}

Letting $A = \exp(a)$, we can rewrite Equations \ref{eq:gompertz} and \ref{eq:MoranRicker}, the Gompertz and Moran-Ricker process equations (respectively), in the form:
\begin{align}
    X_t &= A (X_{t-1})^{b+1} \exp(\epsilon_t),\text{ } \epsilon_t \sim \mathcal{N}(0, \phi) \label{eq:gompertz_re} \\
    X_t &= A X_{t-1} \exp(b X_{t-1}) \exp(\epsilon_t),\text{ } \epsilon_t \sim \mathcal{N}(0, \phi) \label{eq:MoanRicker_re}
\end{align}
More generally, we can think of both the Gompertz and Moran-Ricker process models as belonging to a class of lognormally distributed process models that have the form:
\begin{align}
    &X_{t} = f^* (X_{t-1} \lvert \Theta) \exp(\epsilon_t), \text{ } \epsilon_t \sim \mathcal{N}(0, \phi),\text{ } f^* (X_{t-1} \lvert \Theta) > 0,
    \label{eq:biased_process}
\end{align}
where $f^* (X_{t-1} \lvert \theta)$ is the model of choice for the non-negative physical process, and the process error is governed by a multiplicative zero mean lognormal distribution. Key statistical properties (conditional mean, conditional variance, and conditional median) of this class of process models are:
\begin{align}
\mathbb{E}[X_t \lvert X_{t-1}] &= f^* (X_{t-1} \lvert \Theta) \exp ((2\phi)^{-1}), \label{eq:general_mean_ln} \\
\mathbb{V}[X_t \lvert X_{t-1}] &= f^* (X_{t-1} \lvert \Theta)^2 \exp(\phi^{-1}), \Big( \exp(\phi^{-1}) - 1 \Big) \\
\mathscr{M}[X_t \lvert X_{t-1}] &= f^* (X_{t-1} \lvert \Theta).
\end{align}
When looking at the conditional expected value, we see that this class of models is biased in terms of mean process evolution, but unbiased in terms of median process evolution. Further, the variance of the latent state is controlled by the value of the process model $f^*(\cdot)$, with larger values of the process assumed to have larger variation than smaller ones. Therefore, in the class of process models that describe the Gompertz and Moran-Ricker models, the density dependent relationship of the latent process variance is assumed to scale quadratically with the predicted value of the process model. 

The generalization of the Gompertz and Moran-Ricker process models also holds true for the observation model. For example, for continuous responses $Y$, the assumed relationship between an arbitrary observation $Y_t$ and the corresponding latent state $X_t$ for the Gompertz model may be given by:
\begin{align}
    Y_t = X_t \exp(\epsilon_{obs,t}), \text{ } \epsilon_{obs,t} \sim \mathcal{N}(0, \tau)
    \label{eq: obs_model}
\end{align}
Broadly, the Gompertz and Moran-Ricker observation models belong to the same class of lognormally distributed observation models in Equation \ref{eq: obs_model} that have the form:
\begin{align}
    &Y_{t} = g^* (X_{t} \lvert \Theta) \exp(\epsilon_{obs, t}), \text{ } \epsilon_{obs, t} \sim \mathcal{N}(0, \tau),\text{ } g^* (X_{t} \lvert \Theta) > 0,
    \label{eq:biased_obs}
\end{align}
where $g^* (X_t \lvert \Theta)$ is now interpreted as the model used to link our observations to their respective latent states, and the measurement error is dictated by a multiplicative zero mean lognormal distribution. The conditional mean, conditional variance, and conditional median of this class of observation models are given by:
\begin{align*}
\mathbb{E}[Y_t \lvert X_{t}] &= g^* (X_{t} \lvert \Theta) \exp ((2\tau)^{-1}), \\
\mathbb{V}[Y_t \lvert X_{t}] &= g^* (X_{t} \lvert \Theta)^2 \exp(\tau^{-1}) \Big( \exp(\tau^{-1}) - 1 \Big), \\
\mathscr{M}[Y_t \lvert X_{t}] &= g^* (X_{t} \lvert \Theta).
\end{align*}
 Importantly, while this class of observation models assumes that observations are unbiased in log space, they assume there is systematic observation bias in their measurement process, which is given by $B(Y_t \lvert X_t) = g^* (X_{t} \lvert \Theta) (1 - \exp ((2\tau)^{-1}))$.

Though we have only examined the Gompertz and Moran-Ricker models when discussing the class of lognormal process and observation models presented here, there are many examples of lognormal modeling frameworks in the animal population modeling literature \citep[see][for examples]{BUCKLAND2004, Knape2011} and fisheries modeling  literature \citep[see][for examples]{MAUNDER2015, samu2015} that fall into the classes we discuss here. Including a term to correct the bias is a common method used in applications to rectify assumptions of biased process evolution and biased observations \citep{Knape2011, MAUNDER2015, samu2015} but this bias correction term changes the variance structure, as the lognormal variance is a function of both $\mu$ and $\sigma^2$.

\subsection{Lognormal Moment Matching Models}
Instead of assuming an unbiased median process evolution and a density dependent variance structure, a modeler developing a lognormal SSM for their application may want to specify the mean evolution and the variance structure of the stochastic lognormal process model. That is, we desire a lognormally distributed stochastic evolution function such that $\mathbb{E}[X_t \lvert X_{t-1}] = f^*(X_{t-1} \lvert \Theta)$ and $\mathbb{V}[X_t \lvert X_{t-1}] = \phi_t ^{-1}$. 
We can create a {\it process} model with these characteristics by using a moment matching transformation on the mean and precision, specifically: 
\begin{align}
\mu^* _t &= \log \Big( \frac{f^*(X_{t-1} \lvert \Theta)^2}{\sqrt{f^*(X_{t-1} \lvert \Theta) ^2 + \phi _t^{-1}}} \Big), \label{eq:lnm3_process_mean} \\
\phi^* _t &= \log \Big(1 + (f^*(X_{t-1} \lvert \Theta) ^2 \phi_t)^{-1} \Big)^{-1}. \label{eq:lnm3_process_var}
\end{align}
This approach provides a stochastic lognormal process model with the desired properties (see Appendix \ref{appen_derivation} for derivation):  
\begin{align*}
X_t \lvert X_{t-1} &\sim \text{Lognormal} (\mu^* _t, \phi^* _t), \\
\mathbb{E}[X_t \lvert X_{t-1}] &= f^*(X_{t-1} \lvert \Theta), \\
\mathbb{V}[X_t \lvert X_{t-1}] &= \phi_t ^{-1}.
\end{align*}
Thus, there is a framework for embedding the process model $f^*(X_{t-1} \lvert \Theta)$ such that the temporal evolution is unbiased, and that allows for a flexible way to model the variance of the process through time. For example, we can choose $\phi_t = \phi$ to model constant variance through time, $\phi_t = \phi / f^* (X_{t-1} \lvert \Theta)^2$ to model density dependent variance, or $\phi_t = \phi \exp(-t^{-1})$ to model variance that dissipates over time. 

The lognormal moment matching approach can be similarly applied to the {\it observation} model to produce a lognormally distributed observation density with the properties $\mathbb{E}[Y_t \lvert X_t] = g^*(X_{t} \lvert \Theta)$, $\mathbb{V}[Y_t \lvert X_t] = \tau_t$ by using the transformation:
\begin{align}
\mu^* _{t,obs} & = \log \Big( \frac{g^*(X_{t} \lvert \Theta)^2}{\sqrt{g^*(X_{t} \lvert \Theta) ^2 + \tau _t^{-1}}} \Big), \label{eq:lnm3_obs_mean}\\
\tau^* _t &= \log \Big(1 + (g^*(X_{t} \lvert \Theta) ^2 \tau_t)^{-1} \Big)^{-1}. \label{eq:lnm3_obs_var}
\end{align}
With the forms of both the process model and observation model fully specified in this fashion, it is possible to write a likelihood equation for a model that includes both components, that we call the Lognormal Moment Matching model (LNM3). Suppose that we have observations $\mathbf{Y}$ at a subset of time points $I \subset \{ 1, \dots, T \}$. Then the likelihood for the LNM3 is given by
\begin{align}
&\mathscr{L}(X_{1:T}, \Theta \lvert \mathbf{Y}) = \nonumber \\ &\prod_{t = 1}^T \text{log}\mathcal{N} \left(\log \left( \frac{f^*(X_{t-1} \lvert \Theta)^2}{\sqrt{f^*(X_{t-1} \lvert \Theta) ^2 + \phi _t^{-1}}} \right), \log \left(1 + (f^*(X_{t-1} \lvert \Theta) ^2 \phi_t)^{-1} \right)^{-1} \right) \nonumber \\
&  \times \prod_{i \in I} \text{log}\mathcal{N} \left(\log \left( \frac{g^*(X_{i} \lvert \Theta)^2}{\sqrt{g^*(X_{i} \lvert \Theta) ^2 + \tau _i^{-1}}} \right), \log \left(1 + (g^*(X_{i} \lvert \Theta) ^2 \tau_i)^{-1} \right)^{-1}\right)
\label{eq:LNMM_ll}
\end{align}

The moment matching method can be used to embed any positive process or observation model and any variance structure into a stochastic lognormal model, and thus helps to maintain biophysical realism when modeling positive processes. Fitting these models as state space models further allows them to easily handle missing data and partition between process and measurement error. Thus the LNM3 approach provides a framework that is flexible, biophysically realistic, and statistically coherent. 

\subsection{Model Fitting}
Our analysis focused on comparing and contrasting the LNM3 approach to the Gompertz and Moran-Ricker approaches, two common lognormal SSMs. We estimate the latent states, precisions, and model parameters for six different models presented here using Bayesian state space models (SSMs) \citep{Hamilton, DurbinKoopman, Petris}. The six different models include the Gompertz SSM, the Moran-Ricker SSM, and four different SSM formulations using the Lognormal Moment Matching (LNM3) method presented in Equations \ref{eq:lnm3_process_mean} -- \ref{eq:lnm3_obs_var}. Model parameters, latent states, and precisions were estimated using Markov Chain Monte Carlo (MCMC) \citep{CasellaRobert}. MCMC is an estimation method that generates samples of parameters from their posterior distributions using Markov chains. Parameter uncertainty can be quantified using posterior samples from these Markov chains.  All models were fit using \texttt{JAGS} \citep{Plummer03jags:a} using the \texttt{rjags} package \citep{rjags} in \texttt{R} version 4.1.0 \citep{R}.

\subsubsection{Gompertz SSM Fitting}

We fit the Gompertz model in log space, by taking the logarithm of the latent states $(X_{1:T})$ and observations ($Y_{t \in I}$), $D_t = \log(X_t)$ $F_i = \log(Y_i)$. Under this transformation, the process model and observation model can be written as a Normal Dynamic Linear Model (NDLM) \citep{HarrisonWest}:
\begin{align}
D_t &= a + (1+b) D_{t-1} + \epsilon_t, \text{ } \epsilon_t \sim \mathcal{N}(0, \phi) \\
Y_i &= D_i + \epsilon_{obs,i}, \text{ } \epsilon_{obs,i} \sim \mathcal{N}(0, \tau)
\end{align}
If the log observations ($\mathbf{F}$) are available at a subset of time points, $I \subset \{1, \dots, T \}$, the likelihood for the Gompertz model can be written as:
\begin{align} %
\mathscr{L}(D_{1:T}, a, b, \tau, \phi \lvert \mathbf{F}) \propto &\prod_{t = 1}^T \sqrt{\phi} \exp \left(- \frac{\phi}{2} (D_t - a - (1+b) D_{t-1})^2 \right) \nonumber \\ 
& \times \prod_{i \in I} \sqrt{\tau} \exp \left(- \frac{\tau}{2} (F_i - D_i)^2 \right).\label{eq: gomplikelihood}
\end{align}

Based on the likelihood for the Gompertz SSM, we need prior distributions for $a, b, \phi, \tau$, and the initial latent state $X_0$. Given the interpretation of $a$ and $b$ as the multiplicative constant and the growth rate in the Gompertz model (Equation \ref{eq:gompertz_re}), we used Uniform$(-10,10)$ priors for both $a$ and $b$. There are important considerations when choosing the prior distributions for $\phi$ and $\tau$. Though the likelihood in Equation \ref{eq: gomplikelihood} can exploit conjugacy and use improper Jeffreys priors \citep{JEFFREYS1946} for both $\phi$ and $\tau$, \cite{Gelman2006} shows that the posterior is sensitive to the choice of $\epsilon$ when using a Gamma($\epsilon, \epsilon$) prior in JAGS to emulate the improper Jeffreys prior. Following the advice of \cite{Gelman2006} and \cite{PolsonScott2012}, we use a central half Cauchy prior distribution for the precision parameters. We note that though these studies recommend central half Cauchy priors on variance parameters, under this choice of prior the inverse variance (precision) is also implied to have a central half Cauchy prior (see Appendix \ref{appen_cauchy} for proof). We chose the initial condition prior, $\pi(D_0)$, to be normally distributed, with an initial mean of $\mu_0$, and an initial precision of $\phi_0$. Thus the priors are given by:
\begin{align}
a &\sim \text{Uniform}(-10, 10), \label{eq:gomp_a_prior} \\
b &\sim \text{Uniform}(-10, 10), \label{eq:gomp_b_prior} \\
\phi &\sim \text{HalfCauchy}(\gamma = 100), \label{eq:gomp_phi_prior}  \\
\tau &\sim \text{HalfCauchy}(\gamma = 100), \label{eq:gomp_tau_prior} \\
D_0 &\sim \mathcal{N}(\mu_0, \phi_0). \label{eq:gomp_d0_prior}
\end{align}
The full conditional distributions for $D_{1:T}$ in the Gompertz model are analytically tractable, allowing for Gibbs sampling \citep{GemanAndGeman}. For interior latent states ($k = 1, 2, \dots, T-1$) the Gibbs updates are given by:

\begin{align}
\pi(D_k \lvert \cdot) \sim N\big(&\mu^* = \frac{\phi((1+b)D_{k-1} + a + (1+b)(D_{k+1} - a)) + \tau F_k \mathbbm{1}_{k \in I} }{\phi (1 + (1+b)^2) + \tau \mathbbm{1}_{k \in I}}, \\
& \phi^* = \phi (1 + (1+b)^2) + \tau \mathbbm{1}_{k \in I} \big). \nonumber
\end{align}
The Gibbs updates for the initial latent state and the final latent state are given by
\begin{align}
\pi(D_T \lvert \cdot) &\sim N\left(\frac{\phi(1 + (1+b) D_{T-1} + a) + \tau F_T \mathbbm{1}_{T \in I})}{\phi + \tau \mathbbm{1}_{T \in I} },  \phi + \tau \mathbbm{1}_{T \in I}\right), \\
\pi(D_0 \lvert \cdot) &\sim N\left( \frac{\mu_0 \phi_0 + \phi (1 + b)(D_1 - a)}{\phi(1 + (1+b)^2) + \phi_0}, \phi(1 + (1+b)^2) + \phi_0 \right),
\end{align}
where the function $\mathbbm{1}_{i \in I}$ is an indicator function that is checking if an observation $F_i$ is available at time $i$. 

We ran the MCMC for the Gompertz model for a total of 10,000 iterations, with a burn-in of 2,000 iterations, and an adaptation period (\texttt{n.adapt} in \texttt{JAGS}) of 1,000. 

\subsubsection{Moran-Ricker SSM Fitting}

We also fit the Moran-Ricker model in log space by taking the logarithm of the latent states $(X_{1:T})$ and observations ($Y_{t \in I}$), $D_t = \log(X_t)$, $F_i = \log(Y_i)$. Under this log transformation, the process model and observation model can be written as:
\begin{align}
D_t &= a + D_{t-1} + \exp(b D_{t-1}) + \epsilon_t, \text{ } \epsilon_t \sim \mathcal{N}(0, \phi), \\
Y_i &= D_i + \epsilon_{obs,i}, \text{ } \epsilon_{obs,i} \sim \mathcal{N}(0, \tau).
\end{align}
Assuming that log observations ($\mathbf{F}$) are available at a subset of time points, $I \subset \{1, \dots, T \}$, the likelihood for the Moran-Ricker model fit in log space is given by:
\begin{align}
\mathscr{L}(X_{1:T}, a, b, \tau, \phi \lvert Y_I) \propto &\prod_{t = 1}^T \sqrt{\phi} \exp \left(- \frac{\phi}{2} (D_t - a - D_{t-1} - \exp(b D_{t-1}))^2 \right), \nonumber \\
&\prod_{i \in I} \sqrt{\tau} \exp \left(- \frac{\tau}{2} (F_i - D_i)^2 \right). \label{eq: mr_likelihood} 
\end{align}
Prior distributions for the Moran-Ricker are identical to those chosen for the Gompertz model (Equations \ref{eq:gomp_a_prior} -- \ref{eq:gomp_d0_prior}). $a$ and $b$ were given Uniform($-10, 10$), $\phi$ and $\tau$ were given diffuse half Cauchy priors following recommendations by \cite{Gelman2006} and \cite{PolsonScott2012}, and the initial latent state $D_0$ was given a normal prior centered at an initial mean $\mu_0$ with initial precision $\phi_0$. Altogether, the prior equations are: 
\begin{align*}
a &\sim \text{Uniform}(-10, 10),  \\
b &\sim \text{Uniform}(-10, 10),  \\
\phi &\sim \text{HalfCauchy}(\gamma = 100),   \\
\tau &\sim \text{HalfCauchy}(\gamma = 100),  \\
D_0 &\sim \mathcal{N}(\mu_0, \phi_0). 
\end{align*}

We ran the MC for Moran-Ricker model for a total of 10,000 iterations, with a burn-in of 2,000 iterations and an adaptation period (\texttt{n.adapt} in \texttt{JAGS}) of 1,000. 

\subsubsection{Lognormal Moment Match SSM Fitting}

We explored four different lognormal models using the moment matching approach. These models include: the Gompertz process model embedded for unbiased mean process evolution and unbiased observation model with constant variances (LGC), the Moran-Ricker process model embedded for unbiased mean process evolution and unbiased observation model with constant variances (LMRC), the Gompertz process model embedded for unbiased mean process evolution and unbiased observation model with density dependent variances (LGD), and the Moran-Ricker process model embedded for unbiased mean process evolution and unbiased observation model with density dependent variances (LMRD). For the remainder of this paper, we will be referring to the first two models (LGC, LMRC) as the constant variance models. The term density dependent models will be used to describe the second two models (LGD, LMRD) as well as the classical Gompertz and Moran-Ricker models. We carefully chose these four models to embed common assumptions for different SSM formulations. The unbiased mean process evolution, unbiased observation density function, and constant variance models (LGC, LMRC) were chosen to mimic the assumptions of homoskedastic Gaussian SSMs, which are not frequently used in lognormal SSMs. The embeddings for the LGD and LMRD models were chosen to mimic the variance structure of the Gompertz and Moran-Ricker models while maintaining an unbiased process evolution function and unbiased observation density function.

\begin{table}[h!]
\begin{adjustwidth}{1cm}{1cm}
\begin{center}
 \begin{tabular}{| l | c | c | c| c|} 
 \hline
 Model Name & $f^*(X_{t-1} \lvert \Theta)$ & $\phi_t$ & $g^*(X_i \lvert \Theta)$ & $\tau_i$ \\ 
 \hline
 LGC  & $\exp(a) (X_{t-1})^{b+1}$ & $\phi$ & $X_i$ & $\tau$ \\ 
 \hline
 LMRC & $X_{t-1} \exp(a + b X_{t-1})$ & $\phi$ & $X_i$ & $\tau$  \\
 \hline
 LGD & $\exp(a) (X_{t-1})^{b+1}$ & $\phi(\exp (a) (X_{t-1})^{b+1})^{-2}$ & $X_i$ &  $\tau X_i ^{-2}$ \\
 \hline
 LMRD & $X_{t-1} \exp(a + b X_{t-1})$ & $\phi (X_{t-1} \exp(a + b X_{t-1}))^{-2}$ & $X_i$ & $\tau X_i ^{-2}$ \\
\hline
\end{tabular}
\end{center}
\caption{Process evolution functions, process error structure, observation density function, and observation error structure for the four types of Lognormal Moment Matching SSMs used. LGC and LMRC represent the Gompertz and Moran-Ricker process functions and observation functions with constant process and measurement variance. LGD and LMRD represent the Gompertz and Moran-Ricker process functions and observation functions with density dependent process and measurement variance. }
\label{tb:lnmm}
\end{adjustwidth}
\end{table}

The likelihood for the constant variance models is obtained by substituting the values for $f^*(X_{t-1} \lvert \Theta)$, $\phi_t, g^*(X_i \lvert \Theta)$ and $\tau_i$ from Table \ref{tb:lnmm} into the LNM3 likelihood (Equation \ref{eq:LNMM_ll}). For the constant variance models, two of the prior choices were modified from the priors used for the Gompertz and Moran-Ricker models. First, the prior distribution on $a$ was modified to a strictly positive uniform distribution, to reflect the non-negativity of the latent process $X_{1:T}$. Second, the  prior on the initial condition, $X_0$, was changed from $\mathcal{N}(\mu_0, \phi_0)$ to $\text{Log}\mathcal{N}(\mu_0, \phi_0)$ since the constant variance models are not being fit in log space. Priors for $b, \phi,$ and $\tau$ (Equations \ref{eq:gomp_b_prior} - \ref{eq:gomp_tau_prior}) were not changed. Altogether, the priors for the constant variance models are
\begin{align}
a &\sim \text{Uniform}(0, 10), \\
b &\sim \text{Uniform}(-10, 10), \\
\phi &\sim \text{HalfCauchy}(\gamma = 100), \\
\tau &\sim \text{HalfCauchy}(\gamma = 100), \\
D_0 &\sim \text{Lognormal}(\mu_0, \phi_0).
\end{align}

The density dependent moment matching models were fit in log space, by taking the log of the latent states and observations; $D_t = \log(X_t),  F_i = \log(Y_i)$. The model likelihood in log space for the LGD model can be written as:
\begin{align}
\mathscr{L}(D_{1:T}, a, &b, \tau, \phi \lvert \mathbf{F}) \propto \nonumber \\ &\prod_{t = 1}^T \frac{\exp \left(- \frac{\log(1 + \phi^{-1})^{-1}}{2} (D_t - (a + (1+b) D_{t-1} - .5 \log(1 + \phi^{-1})))^2 \right)}{\sqrt{\log(1 + \phi^{-1})}} \nonumber \\
&\prod_{i \in I} \frac{\exp \left(- \frac{\log(1 + \tau^{-1})^{-1}}{2} (F_i - (D_i - .5 \log(1 + \tau^{-1})))^2 \right)}{\sqrt{\log(1 +\tau^{-1})}}. \label{eq:LGD_likelihood}
\end{align}
Similarly, the model likelihood in log space for the LMRD model can be written as:
\begin{align}
\mathscr{L}(&D_{1:T}, a, b, \tau, \phi \lvert \mathbf{F}) \propto \nonumber \\ &\prod_{t = 1}^T \frac{\exp \left(- \frac{\log(1 + \phi^{-1})^{-1}}{2} (D_t - (a + D_{t-1} + b\exp(D_{t-1}) - .5 \log(1 + \phi^{-1})))^2 \right)}{\sqrt{\log(1 + \phi^{-1})}} \nonumber \\
&\prod_{i \in I} \frac{\exp \left(- \frac{\log(1 + \tau^{-1})^{-1}}{2} (F_i - (D_i - .5 \log(1 + \tau^{-1})))^2 \right)}{\sqrt{\log(1 +\tau^{-1})}}. \label{eq:LMRD_likelihood}
\end{align}
Priors for these two density dependent moment matching models were chosen to be identical to those chosen for the Gompertz and Moran-Ricker models (Equations \ref{eq:gomp_a_prior} -- \ref{eq:gomp_d0_prior}).

The specification of the density dependent variance as $\phi_t = \phi f^*(X_{t-1} \lvert \Theta)^2$ and the log-linear nature of the Gompertz curve conveniently lead to closed form full conditional distributions for the latent states in the LGD model. Similar to the Gompertz model, this allows for Gibbs sampling to be used to update the latent states, which helps to decrease computation time. The derivation of the full conditional distributions is involved, and can be found in the Appendix.

We ran the MC for each of the four moment matching models for a total of 10,000 iterations, with a burn-in period of 2,000 iterations, and an adaptation period (\texttt{n.adapt} in \texttt{JAGS}) of 1,000.

\subsection{Simulation Study}

We designed a simulation study to assess the forecasting performance of the six models presented here, both under the cases where they are the true generating models and where the models are mis-specified. Our three primary objectives for the simulation study were: 1) assess forecasting performance of each model when it is the true generating model and when it is misspecified, when observation precision is being estimated 2) assess forecasting performance of each model when it is the true generating model and when it is misspecified, with fixed observation precision; 3) analyze the estimability of precisions for the models considered in this manuscript. 

To assess the forecast performance of our models, we used proper scoring rules \citep{strictly_proper}. Broadly, proper scoring rules use information about the predictive distribution coupled with observations to assign a measure of agreement of the forecast and the observations \citep{kruger2021}. Specifically, \citet{strictly_proper} define a scoring rule to be proper if the expected value of the score is maximized by a draw from the true forecast distribution, and show that both the Continuous Ranked Probability Score  \citep[CRPS;][]{matheson_winkler1976} and the Ignorance \citep[IGN;][]{good1952, EvaluatingProbabilisticForecastsUsingInformationTheory} score are proper scoring rules.  Given a probability density function $f(\cdot)$ and corresponding cumulative distribution function $F(\cdot)$ for our forecast, and an observation $y$, the IGN and CRPS may be written:
\begin{align}
\text{IGN}(y) &= -\log(f(y));  \\
\text{CRPS}(y) &= \int_{\mathbb{R}} (F(z) - \mathbbm{1}_{z \geq y} )^2 dz .
\end{align}
For our simulation studies, we consider both the CRPS and IGN scores, so that we have one scoring rule defined in terms of the probability density function (IGN) and one scoring rule defined in terms of the cumulative distribution function (CRPS). We use the IGN and CRPS scores to quantitatively compare forecasts, with lower scores within a scoring rule indicating a better performance. CRPS has support over the positive real line, $[0, \infty)$, while IGN can take values between $[-\log(f(y^*)), \infty)$, where $y^* = \text{argmax}_{y \in \mathbb{R}} f(y)$.

To analyze the estimability of precision parameters in each of our six models, we used the empirical coverage rate of the 95\% highest posterior density credible intervals for the precision parameters. \cite{augermethe2016} show that SSMs can have difficulty recovering the process and observation precisions even when the models are linear and Gaussian. To perform a thorough analysis on the estimability of the precision parameters, we fit each model under two different scenarios. In the first scenario, we fixed the observation precision and estimated the process precision for each model. In the second scenario, we estimated both the observation precision and the process precision for each model. This approach allowed us to assess the estimability by looking at the increase in empirical coverage rate for the process precision when the observation precision is fixed compared to when the observation precision is estimated. By choosing to use coverage and proper scoring rules to quantify our model performance we follow a common approach used in the literature -- using frequentist concepts to assess Bayesian models \citep{box1980, rubin1984, Little2006, Little2012}.
\begin{table}[h]
\begin{adjustwidth}{1cm}{1cm}
\begin{center}
 \begin{tabular}{| c | c | c | c| c| c| c|} 
 \hline
 Param. & Gomp & MR & LGC & LMRC & LGD & LMRD \\ 
 \hline
 $\exp(a)$  & .82& 1.26 & 1.21 & 1.11 & 1.21 & 1.11  \\ 
 \hline
 $b$ &  -.658 & -.034 & -.099 & -.014 & -.099 & -.014  \\
 \hline
 $\phi$ & 70.2 & 51.9 & 4 &  4 & 70.2 & 70.2 \\
 \hline
 $\tau$  & 188.7 & 188.7 & 4 & 4 & 188.7 & 188.7 \\
\hline
\end{tabular}
\end{center}
\caption{Parameter values used for the six different model formulations to create thirty synthetic datasets for each generating model. }
\label{tb:simvals}
\end{adjustwidth}
\end{table}

To quantify our three objectives, we performed the simulation study as follows: for each of the six models, thirty different synthetic datasets of length 575 were generated by simulating from the underlying process model and observation model, with parameter values taken from Table \ref{tb:simvals}. Parameter values for each model were chosen so that the systems had similar mean dynamics for each generating model. Each synthetic dataset was fit to each of the six models discussed here. Models were initially fit with the first 365 days of data, and then forecasts were computed for days 366 -- 372, a seven day forecast horizon. The average IGN and CRPS for the seven day forecasts horizons were computed using the \texttt{logs\_sample} and \texttt{crps\_sample} function from the \texttt{scoringRules} {\sf R} package \citep{scoringRulespackage}. We also saved the highest posterior density credible intervals for the estimated precisions. We then re-fit the models with the first 372 days of data, and forecasts were computed for days 373--379. This process was repeated until the synthetic data was exhausted. Overall, each individual synthetic data set was fit a total of 12 times -- once by each model with observation precision known and once by each model with observation precision estimated, for a total of 2,160 simulations. 

This simulation study design helps us to assess our three objectives. Our first two objectives were to assess the forecasting performance of each model under model mis-specification; both in the case of estimated observation precision and in the case of fixed observation precision. Our simulation study had each model fit each dataset twice: once with observation precision fixed and once with observation precision estimated. The forecasts for each precision scenario were evaluated using the CRPS and IGN scores. By using the same synthetic datasets, we were able to isolate the differences in forecast performance for each precision scenario and quantify the difference using paired t-tests with a Holm-Bonferroni adjustment \citep{Holm1979ASS}. With thirty different synthetic datasets generated by each model and each synthetic dataset evaluated over thirty forecast horizons by each of the six models, we were able to get a comprehensive picture of how each model performed under model mis-specification for each of the precision scenarios. Our third objective was to analyze the estimability of precisions for the models considered in this manuscript. We used differences between the scores of the forecasts and coverage rates of the precision credible intervals to determine the severity of underlying precision estimation problems. In the absence of estimation problems, we expect that the empirical coverage rates of the precision parameters should achieve close to the nominal coverage rate of 95\%. CRPS and IGN are proper scoring rules, and their expected scores are maximized by draws from the true forecast distribution \citep{strictly_proper}. Thus we expect that across a large number of simulations, the true generating model should score best in terms of CRPS and IGN in the absence of estimation problems.

\subsection{Application: Leaf Area Index Predictions}

We examined the differences in predictive performance between two different LN-SSM formulations by modeling Leaf Area Index (LAI) at University of Notre Dame Environmental Research Center (UNDE), a National Ecological Observatory Network (NEON) site. LAI is the total surface of leaves per area of ground and is a metric of photosynthetic capacity of an ecosystem. We obtained estimates LAI at UNDE from Oak Ridge National Laboratory Distributed Active Archive Center using their fixed subsets feature \citep{ornlsubset:23Mar2022_16:54:48_545152363}. The dataset contains estimates of LAI at four day intervals, computed using reflectance estimates the from Moderate Resolution Imaging Spectroradiometers (MODIS) satellite as inputs to a phenological model that predicts LAI \citep{modis}. The two statistical models that we considered were a biased LN-SSM, based on Equations \ref{eq:biased_process} and \ref{eq:biased_obs}, and a moment matching LN-SSM, based on Equations \ref{eq:lnm3_process_mean}, \ref{eq:lnm3_process_var}, \ref{eq:lnm3_obs_mean}, and \ref{eq:lnm3_obs_var}. Our analysis was designed with two primary questions in mind: 1) can we construct informed out-of-sample predictions for LAI for long horizons (e.g., multiple seasons) by embedding a process-based ecosystem model into a lognormal state space framework?; 2) do the moment matching and biased formulations show differences in out-of-sample predictive performance for long horizons, when measured by CRPS and IGN?

To fit the LAI data, we used a reduced version of the DALEC2 process-based ecosystem model \citep{BLOOM}. In DALEC2, the LAI is modeled as a function of the carbon stored in foliage. Rather than fitting the full model (a six dimensional dynamical system with 23 process parameters) we only considered the interaction between foliage carbon ($C_f$) and the labile carbon ($C_{lab}$). This reduced the model to a two dimensional dynamical system with eight process parameters. The form of the reduced process model is given as:
\begin{align}
    &\mathbf{C}^{(t)} = M_t \mathbf{C}^{(t-1)} + p_t, \text{ where} \label{eq:dalec2_r} \\
    &M_t = \begin{bmatrix}
    1 - \Phi_{f} ^{(t)} & \Phi_o^{(t)}  \\
     0 & 1 - \Phi_o^{(t)}
  \end{bmatrix},
  \mathbf{C}^{(t-1)} = \begin{bmatrix}
      C_f ^{(t-1)} \\
      C_{lab} ^{(t-1)}
     \end{bmatrix}, p_t = \begin{bmatrix}
           G(\mathbf{D^{(t)}}, c_{lma}, c_{eff}) f_f \\
           G(\mathbf{D^{(t)}}, c_{lma}, c_{eff}) f_{lab}
         \end{bmatrix} \nonumber \\
         &\text{LAI} ^{(t)} = \frac{C_f ^{(t)}}{c_{lma}} \label{lai_obs}
 \end{align}
\begin{align}
&\Phi_f(t, d_{f}, c_{r}, c_{lf}) = \sqrt{\frac{2}{\pi}} \cdot  \frac{-\log(1 - c_{lf})}{c_{rf}} \exp \left(- \left(\sin\left( \frac{t - d_{f} + \psi_f }{s} \right) \frac{\sqrt{2}s}{c_{rf}} \right)^2 \right), \\
&\Phi_o(t, d_{o}, c_{ro}) = \sqrt{\frac{2}{\pi}} \left ( \frac{6.9088}{c_{ro}} \right) \exp \left(- \left(\sin\left( \frac{t - d_{o} + .6245 c_{ro} }{s} \right) \frac{\sqrt{2}s}{c_{ro}} \right)^2 \right),
\end{align}
where $s = 365.25 / \pi$ and $\psi_f = -\sqrt{W_0\left( \left( 2\pi \log(1 - c_{lf})^2 \right)^{-1}\right)}  / \sqrt{2}$, where $W_0$ is the principal branch of the Lambert W function \citep{lambert}, and $G(\mathbf{D^{(t)}}, c_{lma}, c_{eff})$ is the output of the Aggregated Canopy Model (ACM) for gross photosynthetic production \citep{WILLIAMSACM}. $\mathbf{D}^{(t)}$ represents meteorological driver variables for day $t$ that include: daily minimum and maximum temperatures ($^\circ$C), daily incoming shortwave radiation (gCm$^{-1}$), and atmospheric carbon (CO$_2$ ppm). Minimum temperatures, maximum temperatures, and daily shortwave radiation were obtained from the National Ecological Observatory Network \citep[NEON;][]{NEON_data, neontempdata_DP1.00003.001/provisional, neon_swr_DP1.00022.001/provisional}. We imputed any missing NEON observations using a piece-wise linear interpolation. We took monthly measurements  of atmospheric carbon from the Scripps Project \citep{Keeling2005}, and interpolated daily measurements by assigning the monthly values to each day within the month. 

To fit the reduced DALEC2 as a LN-SSM, we embedded the process-based model (Equation \ref{eq:dalec2_r}) as the state component and the LAI equation (Equation \ref{lai_obs}) as the observation component. To model the variance structure of the process evolution function and observation function, we used a density dependent variance. We chose a density dependent variance structure because we expect more process variation and measurement error for foliage carbon when it is large during the early spring and summer months, and less when it is low in the winter months. We considered two different LN-SSM formulations with density dependent variance structures. The first model, the biased model, embeds the process and observation components in a biased manner using Equation \ref{eq:biased_process} and Equation \ref{eq:biased_obs}. The second model that we consider embeds the process and observation components in an unbiased manner using a moment matching approach. The process evolution function and the observation function for the moment matching model are given by:
\begin{align*}
&\mathbf{C}^{(t)} \lvert \mathbf{C} ^{(t-1)} \sim \text{MV}\log \mathcal{N} \left( \log\left(M_t \mathbf{C}^{(t-1)} + p ^{(t)}\right) - .5 \log(\mathbf{11}^T + \Omega_m) \mathbf{1}, \log(\mathbf{11}^T + \Omega_m)  \right), \\
&\text{LAI}^{(i)} \lvert \mathbf{C}^{(i)} \sim \log \mathcal{N} \left( \log\left(\frac{C_f ^{(i)}}{c_{lma}}\right ) - .5 \log(1 + \tau_m ^{-1}), \log(1 + \tau_m^{-1})  \right),
\end{align*}
where $M_t, p_t$ and $\mathbf{C^{(t)}}$ take on the same values from Equation \ref{eq:dalec2_r}, $\mathbf{1}$ represents a $2\times1$ vector of ones, $\Omega_m = \text{Diag}(\omega^2 _{f,m} , \omega^2 _{lab,m})$, and the $\log$ in the process evolution variance represents taking the component-wise logarithm of each element of the matrix/vector.

The process evolution function and observation function for the biased model are given by:
\begin{align*}
&\mathbf{C}^{(t)} \lvert \mathbf{C} ^{(t-1)} \sim \text{MV}\log \mathcal{N} \left( \log\left(M_t \mathbf{C}^{(t-1)} + p ^{(t)}\right) , \Omega_b  \right), \\
&\text{LAI}^{(i)} \lvert \mathbf{C}^{(i)} \sim \log \mathcal{N} \left( \log\left(\frac{C_f ^{(i)}}{c_{lma}}\right ), \tau_b^{-1}  \right),
\end{align*}
where $\Omega_b = \text{Diag}(\omega^2 _{f,b} , \omega^2 _{lab,b})$, and the $\log$ once again represents a component-wise logarithm. The subscripts $b$ and $m$ are used to distinguish between parameters of the biased and moment matching models, respectively. Additional information on expected values and variances for the two different DALEC2 LN-SSM formulations can be found in Table \ref{tb:lai_moments} in the Appendix. 

The equations above that we use to define our LN-SSM also highlight an interesting difficulty of our analysis: we are fitting a two state dynamical system, but we only have observations for one of the states, ($C_f$). Although we are primarily interested in predicting LAI, which is a function of $C_f$, we still need to estimate labile carbon so that we can model it's contribution to foliage during the period of leaf regrowth. This also highlights an advantage of our state space modeling approach: the uncertainty from our lack of labile carbon measurements is propagated through time, giving us a more complete picture of the uncertainty in the foliage carbon.

We treated two model parameters as fixed: the Leaf Mass per Area ($c_{lma}$) and the density dependent observation precision parameter ($\tau$). We fixed $c_{lma}$ to avoid identifiability issues with the canopy efficiency parameter ($c_{eff}$), as the two parameters appear exclusively together in the ACM \citep{WILLIAMSACM}. For both models, we fixed $c_{lma}$ to a value of 75, based on empirical results from \cite{serbin} that were calibrated specifically at UNDE. We estimated the density dependent observation precision parameter using historical data from MODIS, which reports both the standard deviation and the mean of the LAI estimates. For the moment matching model, we took the median value of the means divided by the standard deviations. This gave us a value of $\hat{\tau}_m \approx 4$. For the standard lognormal SSM formulation, we used the form of the variance for the lognormal distribution (Equation \ref{eq:general_mean_ln}) to obtain $\hat{\tau}_m \approx 4.18$. For the DALEC2 process parameters, we used uniform prior distributions over the range of acceptable values taken from \cite{BLOOM}. For the density dependent process variance components, we used a $\text{Uniform}(0,1)$ distribution as the prior. We chose this because the $\omega$ parameters for the biased and moment matching parameters have the interpretation that they roughly control the average proportion of process error at each time point.  All prior distributions for parameters were identical for the moment matching model and the biased model. A full table detailing prior distributions and fixed parameters can be found in Appendix \ref{appen_d2}.

We fit both of the models using particle MCMC \citep[pMCMC;][]{pMCMCAndrieu}. We ran each model for a total of 881 days, from September 5th, 2019 to February 2nd, 2022. We used four day MODIS LAI measurements from September 5th, 2019 to September 6th, 2021 to fit each model, and we used the remaining MODIS LAI measurements from September 7th, 2021 to February 2nd, 2022 to assess out of sample prediction. To assess out of sample predictions, we used the pMCMC samples of the latent states to generate samples from the posterior predictive distribution for the observations. We then used these samples to validate against the out of sample MODIS LAI measurements using CRPS and IGN  using the \texttt{scoringRules} package \citep{scoringRulespackage} in the \texttt{R} programming language \citep{R}. By doing this, we are scoring on $Y \lvert X$ rather than directly on the observation $Y$, and acknowledge that the LAI observations that we are using for validation have measurement error \citep{ferro_imperfect, bessac_naveau}.   We implemented pMCMC using the \texttt{R} package \texttt{pomp} \citep{pomp_jss}. We ran each model for a total of 100,000 iterations, with a burn in of 50,000 iterations, 500 particles, and an adaptive multivariate normal proposal distribution \citep{andrieu_adaptive, rosenthal} that began using a scaled empirical covariance matrix after 1,000 samples are accepted. To obtain initial parameter estimates that start in a region of high posterior density, we used a Gaussian process surrogate model optimization \citep{gramacy2020surrogates} using \texttt{TRBO} \citep{eriksson2019scalable}, with the particle filter marginal log-likelihood as the objective function. 

LAI is ubiquitous for forecasting changes in carbon stored in different components of a terrestrial ecosystem under different projections of climate. For our analysis, we chose to predict over multiple seasons, in an attempt to emulate a forecasting scenario of LAI response to predicted changes in climate, such as a particularly hot or cold winter.  Our analysis also serves as an efficacy test for predictive modeling of LAI using a state space framework. While much work has been done on LAI prediction using ecosystem process-based models \citep[see][for examples]{mahowald_lai, ercanli_lai}, to our knowledge there has been little work done on predicting LAI by embedding a mechanistic process-based ecosystem model as the process component of a statistical state space model.

\section{Results}
\subsection{Simulation Study}

The first objective of our simulation study was to assess the forecasting performance of each model under model mis-specification when observation precision is estimated. We expected that the true generating model would perform best on average for its thirty synthetic datasets, using the average IGN and CRPS scores over each seven day forecast horizon. For the case where both the process precision, $\phi$, and the observation precision, $\tau$, were estimated, the Gompertz model (Gomp in Table \ref{tb:tau_est}) had the best forecasting performance among the density dependent models (MR, Gomp, LMRD, LGD) for both CRPS and IGN. The unbiased moment matching analog to the Gompertz, LGD, also had a strong performance, scoring second highest for IGN on all four density dependent variance models, and scoring second highest for CRPS on three out of the four. For the constant variance models (LMRC, LGC), the constant variance Moran-Ricker model (LMRC) had the best forecasting performance for both CRPS and IGN. 

\begin{table}[h]
\begin{adjustwidth}{1cm}{1cm}
\begin{center}
 \begin{tabular}{|l|c|c|c|c|c|c|} 
 \hline
 \multicolumn{7}{|c|}{Generating Model} \\
 \hline
 Model & MR & Gomp & LMRC &  LGC & LMRD & LGD \\ 
 \hline
 \multicolumn{7}{|c|}{Average CRPS} \\
 \hline
 MR  & .8587 & .8027 & .6496 & .6362 & .8752 & .8970\\ 
 \hline
 Gomp &  \bf{.8344} & \bf{.7773} & .5997 & .5952 & \bf{.8279} & \bf{.8500} \\
 \hline
 LMRC & \it{.8356} & .7810 & .\bf{5913} & \bf{.5889} & .8411 & .8672  \\
 \hline
 LGC  & .8375 & .7809 & \it{.5927} & \it{.5900} & .8440 & .8672\\
 \hline
 LMRD  & .8380 & .7872 & .6181 & .6100 & .8357 & .8585 \\
 \hline
 LGD  & .8357 & \it{.7789} & .6080 & .6016 & \it{.8310} & \it{.8517} \\
 \hline
 \multicolumn{7}{|c|}{Average IGN Score} \\
 \hline
 MR  & 1.816 & 1.744 & 1.595 & 1.557 & 1.806 & 1.830 \\ 
 \hline
 Gomp &  \bf{1.785} & \bf{1.710} & 1.514 & 1.494 & \bf{1.751} & \bf{1.774} \\
 \hline
 LMRC & 1.812 & 1.757 & \bf{1.456} & \bf{1.452} & 1.858 & 1.915 \\
 \hline
 LGC  & 1.806 & 1.750 & \it{1.458} & \it{1.454} & 1.840 & 1.882 \\
 \hline
 LMRD  & 1.793 & 1.723 & 1.537 & 1.507 & 1.762 & 1.790 \\
 \hline
 LGD  & \it{1.792} & \it{1.714} & 1.530 & 1.500 & \it{1.755} & \it{1.778} \\
 \hline
\end{tabular}
\end{center}
\caption{Average CRPS and IGN scores for the simulations where both $\tau$ and $\phi$ are estimated. Columns represent the generating model for the synthetic datasets and rows represent the models used to fit the datasets. Scores are averaged over thirty different synthetic datasets and thirty different 7 day forecast horizon for each combination of generating model and model used to fit the data. Bolded entries represent the lowest score for a given generating model, and italicized entries represent the second lowest score.}
\label{tb:tau_est}
\end{adjustwidth}
\end{table}

The second objective of our simulation study was to assess the forecasting performance of each model under model mis-specification when observation precision was fixed. For the simulations where the observation precision, $\tau$, was fixed and the process precision, $\phi$, was estimated, we found higher consistency between our scoring rules and the generating model. For the simulations that estimated both precision parameters (Table \ref{tb:tau_est}), the Gomp and LGD models consistently outperformed the other density dependent variance models for both scoring rules. In contrast, the simulations where $\tau$ was fixed and $\phi$ was estimated (Table \ref{tb:notau}) showed a more equal representation among the density dependent models. For the four density dependent generating models, the Moran-Ricker and its unbiased moment matching counterpart scored the best for CRPS, with each model producing the lowest score twice. When measuring performance with IGN, the Moran-Ricker and Gompertz models each scored lowest twice. The density dependent models all scored similarly, agnostic to the choice of true generating model, both for CRPS (max difference $\approx .0052$) and for IGN (max difference $\approx .006$). 

\begin{table}[H]
\begin{adjustwidth}{1cm}{1cm}
\begin{center}
 \begin{tabular}{|l|c|c|c|c|c|c|} 
 \hline
 \multicolumn{7}{|c|}{Generating Model} \\
 \hline
 Model & MR & Gomp & LMRC &  LGC & LMRD & LGD \\ 
 \hline
 \multicolumn{7}{|c|}{Average CRPS} \\
 \hline
 MR  & \it{.8332} & \it{.7770} & .5968 & .5945 & \bf{.8230} & \bf{.8492}\\ 
 \hline
 Gomp &  .8358 & .7774 & .6000 & .5954 & .8276 & .8508 \\
 \hline
 LMRC & .8357 & .7813 & \bf{.5911} & \bf{.5891} & .8337 & .8619  \\
 \hline
 LGC  & .8362 & .7805 & \it{.5929} & \it{.5897} & .8349 & .8603\\
 \hline
 LMRD  & \bf{.8332} & \bf{.7766} & .5971 & .5937 & \it{.8233} & \it{.8495} \\
 \hline
 LGD  & .8359 & .7777 & .5999 & .5959 & .8282 & .8510 \\
 \hline
 \multicolumn{7}{|c|}{Average IGN Score} \\
 \hline
 MR  & \bf{1.785} & 1.711 & 1.502 & 1.487 & \bf{1.748} & \it{1.775} \\ 
 \hline
 Gomp &  1.788 & \bf{1.710} & 1.515 & 1.494 & 1.752 & \bf{1.774} \\
 \hline
 LMRC & 1.819 & 1.764 & \bf{1.455} & \bf{1.453} & 1.848 & 1.905 \\
 \hline
 LGC  & 1.813 & 1.750 & \it{1.458} & \it{1.453} & 1.834 & 1.887 \\
 \hline
 LMRD  & \it{1.786} & 1.712 & 1.505 & 1.486 & \it{1.752} & 1.780 \\
 \hline
 LGD  & 1.788 & \it{1.710} & 1.513 & 1.494 & 1.752 & 1.776 \\
 \hline
\end{tabular}
\end{center}
\caption{Average CRPS and IGN scores for the simulations where $\phi$ is estimated and $\tau$ is known. Columns represent the generating model for the synthetic datasets and rows represent the models used to fit the datasets. Scores are averaged over thirty different synthetic datasets and thirty different 7 day forecast horizon for each combination of generating model and model used to fit the data. Bolded entries represent the lowest score for a given generating model, and italicized entries represent the second lowest score.}
\label{tb:notau}
\end{adjustwidth}
\end{table}

Our third objective was to analyze the estimability of precisions for each of the six models we used. Our first investigation for the estimability of the precision parameters was to quantify the differences in IGN and CRPS scores between the simulations where the observation, $\tau$ was fixed and the simulations where $\tau$ was estimated. This investigation is also intimately related to the performance of the models under model mis-specification, and therefore provides insight for all three of our objectives.  To evaluate the impacts statistically, we used a paired t-test. To do this, we took the average IGN and CRPS scores over each seven day forecast horizon for the two different scenarios, and treating them as "before" and "after". We justify this by noting that each precision scenario was fit using identical synthetic datasets, with the only difference being whether $\tau$ was estimated or not. Unsurprisingly, we found that fixing the observation precision helped to improve forecasting performance for nearly all models. The LGC and LMRC models performed significantly better in terms of CRPS when $\tau$ was fixed (LGC: $p = 4.7\text{e-}05$; LMRC: $p = .0032$), but did not perform significantly better in terms of IGN (LGC: $p = .68$; LMRC: $p = .36$). The Moran-Ricker, LMRD, and LGD models all had statistically significant decreases in IGN (MR: $p < 2\text{e-}16$; LMRD: $p = 2.1\text{e-}09$; LGD: $2.8\text{e-}05$) as well as CRPS (MR: $p < 2\text{e-}16$; LMRD: $p < 2\text{e-}16$; LGD: $1.4\text{e-}05$) when $\tau$ was fixed. Fixing the observation precision did not significantly impact the performance of the Gompertz model for IGN ($p = .84$) or CRPS ($p = .82$). 

Our second investigation for the estimability of the precision parameters was to use the empirical coverage rate of the HPD intervals.  The empirical coverage of the highest posterior density (HPD) credible intervals for $\phi$ unanimously increased for the simulations where $\tau$ was fixed, and all six models had empirical coverage that falls close to the nominal rate. For the simulations where both $\phi$ and $\tau$ are estimated, Table \ref{tb:cvg} shows that the Gompertz model and LGD models produce the best empirical coverage for the precision parameters. This is likely to be related to their excellent performance in these simulations, where other models struggled to consistently produce precision estimates that contained the ground truth in their 95\% HPD intervals. The empirical undercoverage of the Moran-Ricker (MR) model for both $\phi$ and $\tau$ may also explain its poor performance in the forecast results from Table \ref{tb:tau_est}, where it came in last place for CRPS for all six generating models and last place for IGN for three out of six generating models, including the case where it was itself the true generating model. 
\begin{table}[ht]
\begin{adjustwidth}{1cm}{1cm}
\centering
 \begin{tabular}{|l|c|c|c|c|c|c|} 
 \hline
 \multicolumn{7}{|c|}{Generating Model} \\
 \hline
 Coverage & MR & Gomp & LMRC &  LGC & LMRD & LGD \\ 
 \hline
 $\phi$ Cvg  & 72.3\% & 89.1\% & 89.2\% & 86.9\% & 85.8\% & 89\%\\ 
 \hline
 $\tau$ Cvg & 74.7\% & 96.4\% & 92.1\% & 91.8\% & 91.7\% & 95.3\% \\
 \hline
 $\phi$ Cvg, $\tau$ fixed & 95.2\% & 94.8\% & 94.1\% & 93.6\% & 94.2\% & 94.9\% \\
 \hline
\end{tabular}
\caption{Average empirical HPD credible interval coverage for the precisions of each generating model, under the scenarios where both $\phi$ and $\tau$ are estimated and when $\tau$ is fixed. Coverage rates are averaged over all thirty synthetic datasets and all thirty forecast horizons, for a total of 900 samples.}
\label{tb:cvg}
\end{adjustwidth}
\end{table}

\subsection{Leaf Are Index Predictions at UNDE}

\begin{figure}[ht!]
\begin{adjustwidth}{1cm}{1cm}
\centering
\includegraphics[scale = .38]{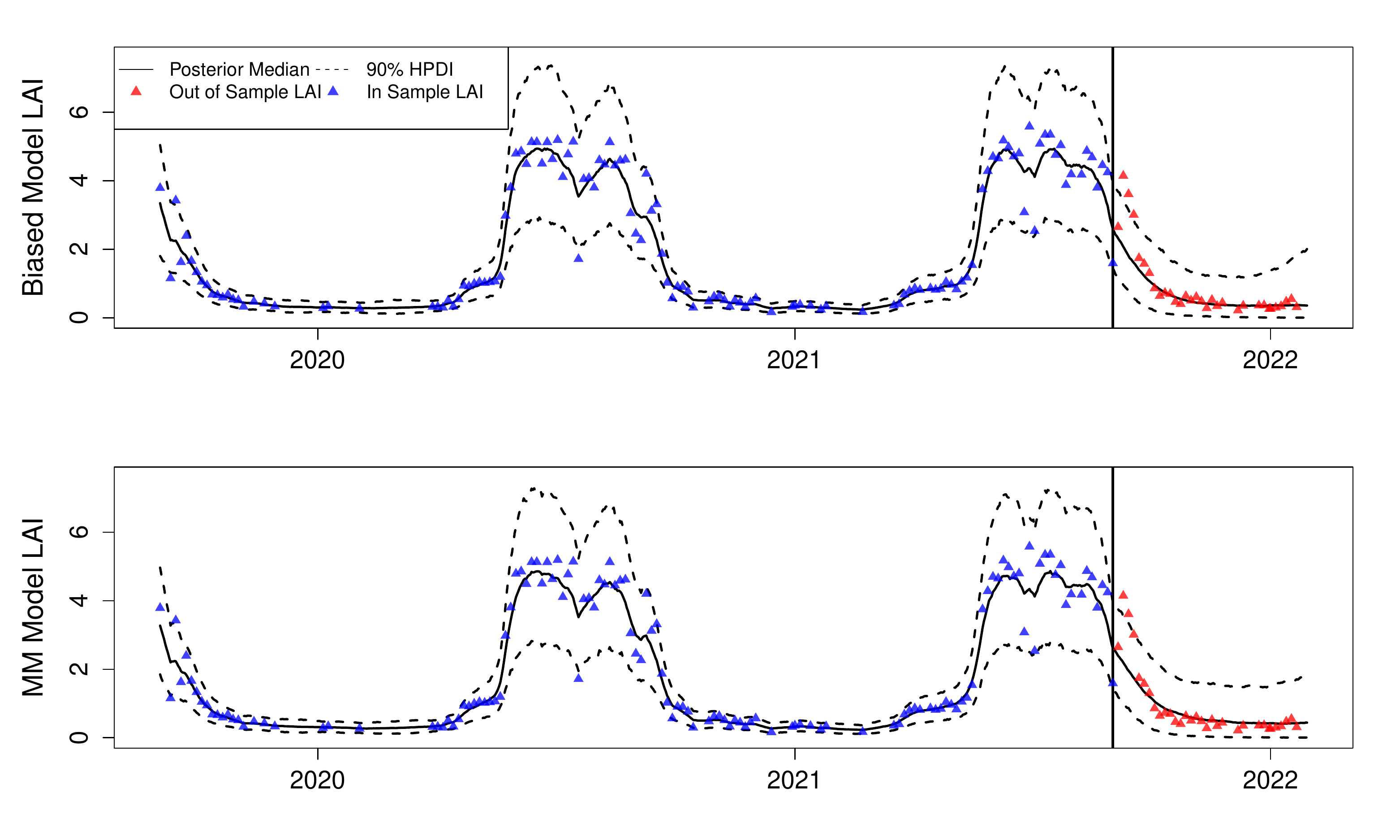}
\caption{Model fits for the biased model (top panel) and moment matching model (bottom panel). Medians (solid line) and 90\% highest posterior density intervals (dashed lines) were computed using 50,000 post burn-in samples of the latent states generated by the pMCMC. Blue triangles denote LAI measurements that were used to train the model, and red triangles denote LAI measurements that were not seen by the model and used only for validation purposes. The vertical black represents the time value where the model began predicting out of sample.}
\label{fig:pred_plot}
\end{adjustwidth}
\end{figure}

We found that both the moment matching LN-SSM and the biased LN-SSM produced predictive distributions that captured the dynamics of both the in sample and the out of sample LAI observations. Both models showed similar fits for the in sample LAI observations when looking at medians (Figure \ref{fig:pred_plot}) and 90\% highest posterior density intervals (Figure \ref{fig:pred_plot}). This was not surprising to us, as both models used an identical process model and prior distributions, and only differed slightly in the formulation of process evolution and observation functions. For the out of sample LAI predictive distributions, the models behaved differently. The biased model (Figure \ref{fig:pred_plot}, top panel) had lower variance at the start of the predictive horizon, and then began to tail off at the end of the horizon. The moment matching model (Figure \ref{fig:pred_plot}, bottom panel) had larger predictive variance at first, but then leveled off and accumulated slowly at the end of the horizon. The median predicted values for the moment matching model are slightly larger than the median predicted values for the biased model over the entire horizon. This is an interesting result: for identical process parameter values, the median of the moment matching models should be strictly smaller than the median of the biased model. This tells us that there is some mismatch between the parameters being estimated for the biased model and the moment matching model. This is something that we must be cautious about, especially since we are using parameters that have physical interpretations and modeling the evolution by embedding an ecosystem model that was designed to be deterministic. 

\begin{figure}[ht!]
\begin{adjustwidth}{1cm}{1cm}
\centering
\includegraphics[scale = .38]{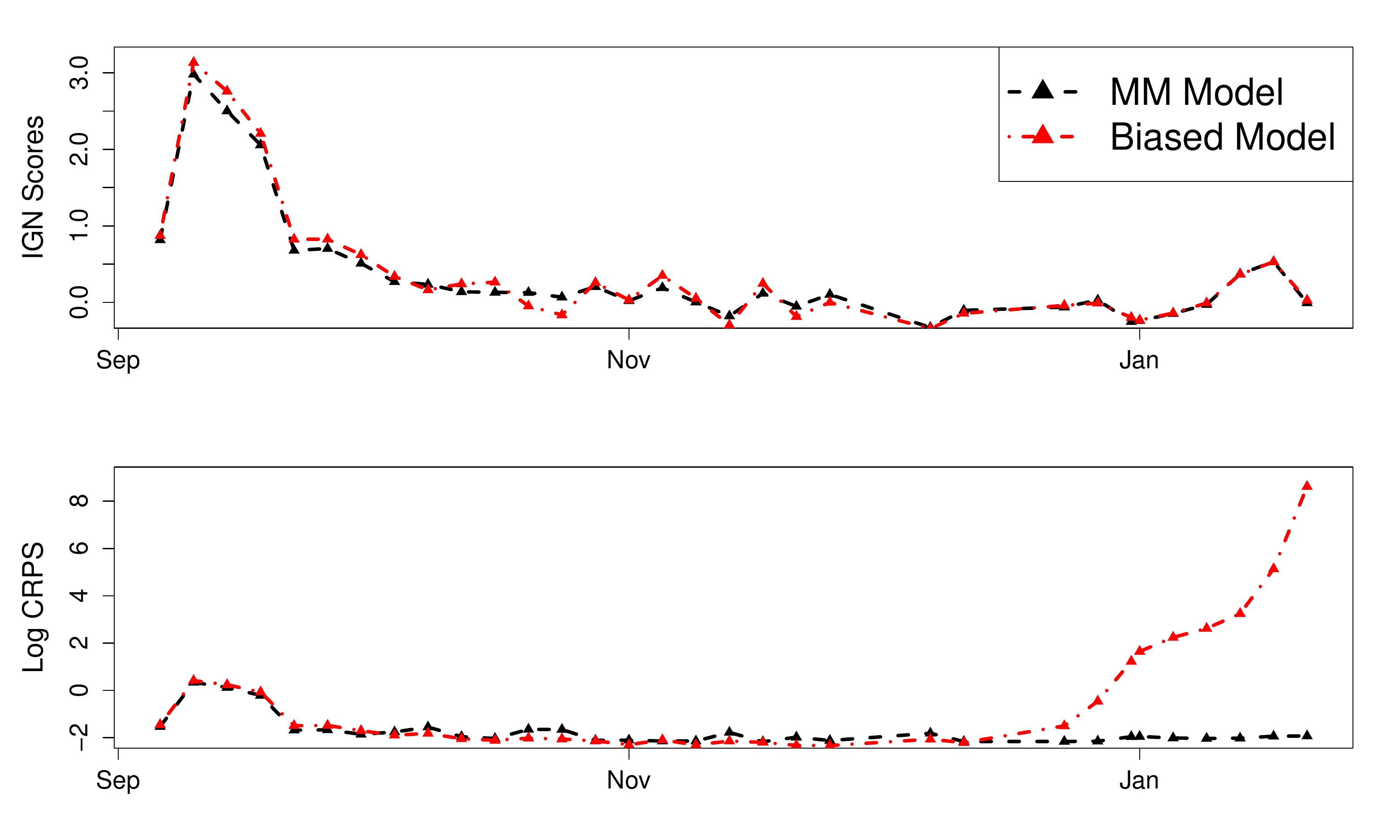}
\caption{Visualizations of the IGN scores (top panel) and the log CRPS (bottom panel). The dashed black line denotes the moment matching model, and the dashed and dotted red line denotes the biased model. Triangles along these lines represent points where there were out of sample LAI measurements for validation. For visualization purposes, we used a linear interpolation between measurements. }
\label{fig:score_plot}
\end{adjustwidth}
\end{figure}

We also found differences in performance between the two models when measured by IGN and CRPS. When measured by mean IGN across the 32 out of sample LAI measurements, the two models had similar performance, with the moment matching model performing slightly better (mean ${\text{IGN}}_{mm} = .3566$; mean ${\text{IGN}}_{biased} = .3850$). The moment matching model had marginally better IGN scores for the observations where neither model performed well (e.g. out of sample measurements two, three, and four, Figure \ref{fig:score_plot}, top panel), and the models had similar IGN score performance otherwise. When measured by mean CRPS across the 32 out of sample LAI measurements, the moment matching model showed a much better performance (mean ${\text{CRPS}}_{mm} = .2389$; mean ${\text{CRPS}}_{biased} = 179.93$). Towards the end of the prediction horizon the CRPS for the biased model quickly increases, while the CRPS for the moment matching model stays comparatively small. We believe that this happens because of an accumulation of bias in the biased model combined with the heavy tails of the lognormal distribution.

\section{Discussion}

Although the Gaussian distribution is a common choice in modeling applications, many ecological processes have strict lower bounds that are not accurately captured by Gaussian models. This mismatch becomes especially problematic in forecasting applications, where uncertainty grows as the forecast horizon increases. However, embedding assumptions about the evolution of the latent process and variance dynamics into non-Gaussian SSM frameworks can be challenging. To remedy this, we proposed a method for embedding non-negative process and observation models with arbitrary variance structures into lognormal Bayesian SSMs using moment matching (LNM3). The primary advantage of our method is flexibility: it allows practitioners to create stochastic lognormal distributions for process and measurement components that are unbiased in terms of their mean evolution and observations,  have a flexible variance that can change through time, and offer a closed form Markov transition density that allows models to be fit with MCMC software such as \texttt{JAGS} \citep{Plummer03jags:a}.

We used a computationally intensive Monte Carlo approach to assess the forecasting performance of the six models discussed here, using a total of 180 synthetic datasets that were fit twelve times each: once by each model with the observation precision estimated, and once by each model with the observation precision fixed. We found that the forecasting performance of our models under mis-specification was heavily dependent on whether or not the observation precision was fixed, and also dependent on the metric used for evaluation: CRPS or IGN. With the observation precision estimated, the Gompertz model had the best average CRPS and IGN scores across all of the synthetic datasets for four out of the six generating models. With the observation precisions fixed at the true values, we found that every model except for the Gompertz model had a significant increase in forecast performance when measured by average CRPS or average IGN. For these simulations, no one model dominated the others in terms of forecast performance for either metric. The Gompertz model outperforming the true generating models identifies one of the difficulties of using proper scoring rules to evaluate forecast performance, especially if using forecasting performance as a way to guide model choice. Although the CRPS and IGN scores should favor the true generating model in expectation, we are unlikely to have access to the ground truth parameter values and instead have to rely on estimates of parameter values from our MCMC.

Even simple linear Gaussian SSMs can be prone to estimation problems, especially for parameters that govern the variance structure of the process and observations \citep{augermethe2016}. We found that our models were no exception to this: when both the process precision and observation precision were estimated, no model came within 5\% of the nominal 95\% coverage rate of the process precision HPDs. We also found that estimation of the precision parameters was closely related to forecasting performance. For the simulation where both observation and process precision parameters were estimated, the two models with the best performance (Gompertz and LGD) were also the models that had empirical coverage rates closest to 95\% for the precision parameters. Similarly, the Moran-Ricker model was 20\% below the nominal coverage rate for both the observation and process precision parameters, and had the worst average CRPS scores for every generating model, including itself. For the simulation studies where the observation precision was fixed, we found that coverage rates for each of the six models were close to the nominal 95\% coverage rate, with the empirical coverage rates ranging from 93.6\% to 95.2\%. This supports the findings from \citet{augermethe2016}, who show that fixing the measurement error in linear Gaussian SSMs can help to alleviate estimation problems. 

We tested the efficacy of our method when applied to a challenging problem by embedding a two-dimensional process-based ecosystem model into a LN-SSM and using it to predict Leaf Area Index (LAI). Overall, we found that both models performed well in reconstructing latent states that had good agreement with the measurements while capturing the dynamics of the out of sample measurements that we used for validation. Both models showed similar fits for the in sample LAI measurements, but showed differences in out of sample predictive performance. The moment matching model, developed using the methodology we present here, had a superior performance for the out of sample LAI when assessed using both IGN scores and CRPS. This lends credence to the idea that having a flexible mechanism for adjusting the process evolution, observation function, and variance structure, even ever so slightly as we did here, can help to better capture out of sample dynamics and improve predictive performance. Towards the end of the predictive horizon, the biased model begins to perform very poorly when measured by CRPS, and the moment matching model continues to perform well. This indicates that the moment matching framework used here may be better for applications that have long predictive horizons, such as multi-year projections of LAI under different climate scenarios. 

Though we saw better performance using our moment matching technique, the analysis that we did here serves mainly as proof-of-concept for state space modeling of LAI, and there is much room for improvement in future work. The largest improvement would be to integrate additional data streams, and to include weather drivers that are forecast. For example, including data on litterfall accumulation from NEON would help to further constrain process parameters, process variance, and potentially improve out of sample prediction performance, and including forecasted weather drivers adds an additional level of uncertainty to our model predictions. Similarly, in future work we can consider additional methods for helping to constrain parameters in the absence of direct observations, such as the ecological data constraints described in \cite{BLOOM}.

Though the methods presented here use the lognormal distribution to represent stochasticity, the moment matching approach  broadly applies to other distributions as well, and provides opportunities for future directions. For example, the gamma distribution has been considered for state space modeling \citep{SmithGamma} and stochastic differential equation modeling \citep{DENNIS1984187} applications, has non-negative support, can be parameterized in terms of its mean and variance using a moment matching approach, and has lighter tail behavior than the lognormal distribution. The beta distribution, which is beginning to be used in SSMs \citep[see][for examples]{osti_1406200, DEO2021104182}, has a useful support for modeling proportions and can also be parameterized in terms of its mean and variance to allow for moment matching approaches. 

In conclusion, to address biological non-realism in models of physical systems, we proposed a novel lognormal state space modeling framework that preserves positivity of the latent process and observations. The methods presented here allow practitioners to embed complex process models and error dynamics into state space models while ensuring that the forecasts they get out of the model agree with the constraints of the system. The flexibility of the moment matching method for representing complex systems along with the variance partitioning of the state space model provide a coherent statistical framework for forecasting, in terms of biophysical realism, forecast assessment, and uncertainty quantification.

\bibliography{sn-article}
\bibliographystyle{apalike}

\section{Appendix}
\subsection{Lognormal Moment Matching} \label{appen_derivation}

Suppose that we are interested in finding a transformation for random variable $X$, $X \sim \text{Lognormal}(\mu^*, \sigma^{*2})$ such that $\mathbb{E}[X] = \mu, \mathbb{V}[X] = \sigma^2$, with pdf given by:

\begin{align*}
f(x \lvert \mu, \sigma) = \frac{1}{\mu \sigma \sqrt{2\pi}} \exp \left( -\frac{(\log(x) - \mu)^2}{2 \sigma^2} \right), \mu \in \mathbb{R}, \sigma >0, x \in (0, \infty)
\end{align*}
Then, we have:
\begin{align*}
&\mathbb{E}[X] = \mu = \exp \left( \mu^* + \frac{\sigma^{*2}}{2} \right) \\
&\mathbb{V}[X] = \sigma^2 =  \left( \exp(\sigma^{*2}) - 1 \right) \exp (2\mu^* + \sigma^{*2})
\end{align*}

Looking at the first equality, we see that $\sigma^{*2} = 2(\log(\mu) - \mu^*)$. We can substitute this into the second equality to get $\mu^*$ written in terms of known constants,
\begin{align*}
\sigma^2 &= \Big( \exp (2 \log(\mu) - 2\mu^*) - 1 \Big) \exp (2\mu^* + 2\log(\mu) - 2\mu^*) \\
&= \Big( \exp (2 \log(\mu) - 2\mu^*) - 1 \Big) \exp (2\log(\mu)) \\
& = (\mu^2 \exp (-2\mu^*) - 1) \mu^2
\end{align*}
Then, rewriting this equation in terms of $\exp(-2\mu^*)$,
\begin{align*}
&\exp(-2\mu^*) = \frac{\sigma^2 + \mu^2}{\mu^4} \\
&-2\mu^* = \log\Big(\frac{\sigma^2 + \mu^2}{\mu^4}\Big) \\
&\mu^* = - \frac{1}{2} \log\Big(\frac{\sigma^2 + \mu^2}{\mu^4}\Big) \\
&\mu^* = \log \Big( \frac{\mu^2}{\sqrt{\mu^2 + \sigma^2}} \Big)
\end{align*}
Substituting our value for $\mu^*$ back into the relation $\sigma^{*2} = 2(\log(\mu) - \mu^*)$, we have:
\begin{align*}
\sigma^{*2} &= 2(\log(\mu) - \mu^*) \\
&= 2 \Big( \log(\mu) - \log\Big( \frac{\mu^2}{\sqrt{\mu^2 + \sigma^2}} \Big) \Big) \\
&= 2 \log \Big( \frac{\sqrt{\mu^2 + \sigma^2}}{\mu} \Big) \\
& = \log \Big( \frac{\mu^2 + \sigma^2}{\mu^2} \Big) \\
&= \log \Big(1 + \frac{\sigma^2}{\mu^2} \Big)
\end{align*}

Thus our desired transform is $\mu^* = \log \Big( \frac{\mu^2}{\sqrt{\mu^2 + \sigma^2}} \Big)$ and  $\sigma^{*2} = \log \Big(1 + \frac{\sigma^2}{\mu^2} \Big)$ 

\subsection{Half Cauchy Precision Prior}\label{appen_cauchy}
Suppose that we are interested in using a half-Cauchy distribution on $\sigma^2$, and want to understand the implied prior on $\phi = \sigma^{-2}$. 
\begin{align}
\sigma^2 \sim \text{HalfCauchy}(\mu = 0, a = \gamma) \\
\pi (\sigma^2) = \frac{2}{\pi \gamma (1 + (\frac{\sigma^2}{\gamma})^2)}, \text{ } \sigma^2 > 0
\end{align}
Let $\phi = \sigma^{-2}$. Then, the pdf of $\phi$ is:
\begin{align}
\pi(\phi) &= \frac{2}{\pi \gamma (1 + (\frac{1}{\phi \gamma})^2)} \frac{1}{\phi^2}, \text{ } \phi > 0 \\
&= \frac{2}{\pi \gamma (\phi^2 + (\frac{1}{\gamma})^2)}  \text{ } \phi > 0 \\
&= \frac{2}{\pi \frac{1}{\gamma} (1 + (\phi \gamma)^2)}  \text{ } \phi > 0
\end{align}
Thus if the prior for $\sigma^2$ is $\sigma^2 \sim \text{HalfCauchy}(\mu = 0, a = \gamma)$, then the implied prior on $\phi$ is $\phi \sim \text{HalfCauchy}(\mu = 0, a = \gamma ^{-1})$

\subsection{Reduced DALEC2 Details} \label{appen_d2}
In this section we provide additional details on the moments and prior distributions for the reduced DALEC2 model that we use \citep{BLOOM}.

\begin{table}[h!]
\begin{adjustwidth}{1cm}{1cm}
\begin{center}
 \begin{tabular}{| l | c | c |} 
 \hline
 Model & MM & Biased   \\ 
 \hline
 $\mathbb{E}[\mathbf{C}^{(t)} \lvert \mathbf{C}^{(t-1)}, \Theta)]$  & $M_t \mathbf{C}^{(t-1)} + p_t$ & $(M_t \mathbf{C}^{(t-1)} + p_t) \exp ((2\phi)^{-1})$ \\ 
 \hline
 $\mathbb{V}[\mathbf{C}^{(t)} \lvert \mathbf{C}^{(t-1)}, \Theta]$ & $\phi^{-1}(M_t \mathbf{C}^{(t-1)} + p_t)$ & $\phi$   \\
 \hline
 $\mathbb{E}[\mathbf{C}^{(i)} _{obs} \lvert \mathbf{C}^{(i)},\Theta]$ & $C_f ^{(i)} c_{lma} ^{-1}$ & $C_f ^{(i)} c_{lma} ^{-1} \exp ((2\tau)^{-1})$ \\
 \hline
 $\mathbb{V}[\mathbf{C}^{(i)} _{obs} \lvert \mathbf{C}^{(i)},\Theta]$ & $\tau^{-1}(C_f ^{(i)} c_{lma} ^{-1})$ & $\phi (X_{t-1} \exp(a + b X_{t-1}))^{-2}$ \\
\hline
\end{tabular}
\end{center}
\caption{Expected values and variances for the process evolution functions and observation functions for the Moment Matching LN-SSM and Biased LN-SSM used to model LAI data at UNDE.}
\label{tb:lai_moments}
\end{adjustwidth}
\end{table}
\begin{table}
\begin{adjustwidth}{1cm}{1cm}
\begin{center}
 \begin{tabular}{|l|l |l| l|} 
 \hline
 Param. & Description &  Units & Prior \\ 
 \hline\hline
 $f_{lab} $ &  Proportion of GPP allocated to labile carbon & unitless & Unif(.01, .5) \\
 \hline
 $f_{f}$  & Proportion of GPP allocated to foliage carbon & unitless & Unif(.01, .5)    \\
 \hline
 $d_o$ & Start day of leaf regrowth onset & unitless & Unif(1, 365)  \\
 \hline
 $d_f$ & Start day of leaf fall & unitless & Unif(1, 365) \\
 \hline
 $c_{eff}$ & Canopy efficiency & unitless & Unif(10, 100) \\
 \hline
 $c_{lf}$ & Proportion of leaves lost annually  & unitless & Unif(.125, 1) \\
 \hline
 $c_{ro}$ & Length of labile carbon release period & day & Unif(10, 100) \\
 \hline
 $c_{rf}$  & Length of leaf fall period & day & Unif(20, 150)  \\
 \hline
 $\omega_{f}$  & Density dependent variance parameter & $g^{.5} \text{C}^{.5} \text{m}^{-1}$ & Unif(0, 1)  \\
 \hline
 $\omega_{lab}$  & Density dependent variance parameter & $g^{.5} \text{C}^{.5} \text{m}^{-1}$ & Unif(0, 1)  \\
 \hline
\end{tabular}
\end{center}
\caption{Information on the 10 parameters estimated in our reduced DALEC2 model \citep{BLOOM}. This information includes notation, interpretations of the parameters, units, and the prior distribution used in our Bayesian Lognormal SSM. Upper and lower bounds for process parameters ($f_{lab}$ through $c_{rf}$) are taken from the table of upper and lower bounds in \cite{BLOOM}. }
\label{table: param_infos}
\end{adjustwidth}
\end{table}

\end{document}